\documentclass[prb,preprint,showpacs,preprintnumbers,amsmath,amssymb]{revtex4}

\usepackage{graphicx}
\usepackage{dcolumn}
\usepackage{bm}
\usepackage{amsmath}

\begin{document}


\title{Plasma Wave Instabilities in Non-Equilibrium Graphene}

\author{Chinta M. Aryal\footnote{Present address: Department of Physics, University of South Florida, 4202 E Fowler Ave, Tampa, FL 33620-7100, USA.} and Ben Yu-Kuang Hu}
\affiliation{Department of Physics, The University of Akron, Akron,OH 44325-4001, USA}

\author{Antti-Pekka Jauho}
\affiliation{Center for Nanostructured Graphene (CNG), Dept.~of Micro and Nanotechnology, Technical University of Denmark, DK-2800 Kongens Lyngby, Denmark}

\date{\today}

\begin{abstract}

We study two-stream instabilities in a non-equilibrium system in which a stream of electrons is injected into doped graphene.  As with equivalent non-equilibrium parabolic band systems, we find that the graphene systems can support unstable charge-density waves whose amplitudes grow with time.  We determine the range of wavevector $\boldsymbol{q}$ that are unstable, and their growth rates.  We find no instability for waves with wavevectors parallel or perpendicular to the direction of the injected carriers.  We find that, within the small wavevector approximation, the angle between  $\boldsymbol{q}$  and the direction of the injected electrons that maximizes the growth rate increases with increasing $\boldsymbol{|q|}$. We compare the range and strength of the instability in graphene to that of two and three dimensional parabolic band systems.

\end{abstract}



\maketitle

\section{\label{sec:level1} Introduction}

Monolayer graphene consists of a single monolayer of carbon atoms arranged in a honeycomb lattice.  There has been a tremendous amount of interest in graphene, both theoretical and experimental, since the initial fabrication by Novoselov \textit{et al.}was reported.\cite{novo2005} Most investigations on the electronic properties of graphene have concentrated on the linear response regime, in which the system is slightly perturbed from its equilibrium state (see, {\it e.g.}, Refs. \onlinecite{RevModPhys.81.109} and \onlinecite{RevModPhys.83.407}).  In this paper, we theoretically investigate a non-equilibrium situation, in which a stream of carriers is injected into doped graphene sample.  We show that this results in an instability in the collective modes of the system, which is analogous to the two-stream instability in classical plasmas.

Under certain conditions in classical plasmas, when there are counter-streaming charged particles, some of the collective modes ({\em i.e.}, charge density waves or plasmons) of the plasma become unstable, in the sense that they initially grow exponentially.  This phenomenon, called the two-stream instability,\cite{chen_intro_plasma} can qualitatively be understood by considering Landau damping, the process in which plasma waves decay in equilibrium.  In equilibrium situations, the interaction between plasma waves and the charged particles in the plasma result in a net transfer of energy from the plasma waves to the individual charged particles, which leads to the decay of the plasma waves.  However, in certain non-equilibrium situations the inverse can occur; the plasma waves absorb a net amount of energy from the charged particles and the waves grow in amplitude.

The possibility of two-stream instabilities in solid-state systems has been studied theoretically by several investigators.  Pines and Schrieffer\cite{PhysRev.124.1387} considered the possibility of these instabilities in 3-dimensional solid-state systems in which both electron and holes are present, such as semimetals or small band-gap semiconductors.  A static electric field would cause the electrons and holes to counter-stream in opposite directions and can in principle produce the instability.   However, in practice, the strong electron-hole scattering in 3-dimensional systems suppresses the counter-streaming motion of the electrons and holes.  Several groups have theoretically investigated two-stream instabilities in coupled two-dimensional structures such as closely-spaced electron and hole doped quantum-wells to separate the oppositely charged carriers.\cite{robi67,krasheninnikov1980instabilities,crow85,hawr86,baks88,PhysRevB.39.6208,PhysRevB.43.14009,PhysRevB.48.9158,PhysRevB.60.8665,Zeba20122309,Gumbs_JAP_2015} However, even in these systems with reduced electron-hole scattering due to the spatial separation between the two carrier species, the strength of electric fields necessary to obtain a sufficient large relative drift velocities of the electron and hole populations would cause heating of the carriers that suppresses the instability.  Another way of obtaining counter-streaming carriers with sufficiently large relative drift velocities is by injection of carriers at high velocities into a doped system.  However, in order to achieve high velocities, for parabolic-band systems, carriers must be injected at high energies.  These high-energy carriers usually scatter inelastically very quickly, typically with optical phonons, which makes it difficult to set a steady-state system with counter-streaming carriers.

In systems with bands that have linear dispersions such as graphene, the velocity of a carrier is independent of the energy.  Thus, it is possible to obtain large relative drift velocities without having to inject carriers with large energies or apply very large electric fields.  Therefore, the reasons given above which tend to suppress the instability in parabolic-band systems do not apply in linear-band systems.   In this paper, we show that it is in fact possible to obtain two-stream instabilities in linear-band systems such as graphene.

The outline of this paper is as follows.  In Section II, we discuss the formalism that we use to calculate the dispersion and growth rates of unstable collective modes ({\em i.e.}, plasmons) both in a doped parabolic band semiconductor (for comparison) and in doped graphene, both with an injected stream of carriers. In Section III, we review the two-stream instability in  two- and three-dimensional parabolic-band systems, and in Section IV, we investigate the two-stream instability in graphene and compare the results to parabolic band systems. Section V contains our discussion and conclusion.

\section{Formalism}
In order to obtain the dispersion relation and growth rates of the plasmons, we analyze the relative dielectric function $\varepsilon(\boldsymbol{q},\omega)$ of extrinsic graphene in the presence of an injected stream of carriers. The dielectric function is defined to be the ratio of an externally induced potential $V_{\mathrm{ext}}(\boldsymbol{q},\omega)$ to the total potential $V_{\mathrm{tot}}(\boldsymbol{q},\omega)$ (the sum of $V_{\mathrm{ext}}(\boldsymbol{q},\omega)$ and the internal potential due to the charge density perturbation in the system),\cite{pines} $\varepsilon(\boldsymbol{q},\omega)$ = $V_{\mathrm{ext}}(\boldsymbol{q},\omega)/ V_{\mathrm{tot}}(\boldsymbol{q},\omega)$. The plasmon modes of wave-vector $\boldsymbol{q}$ are obtained by solving for $\omega_{\boldsymbol{q}}$ in the equation $\varepsilon(\boldsymbol{q},\omega_{\boldsymbol{q}})$ $=0$, since this condition implies the existence of a $V_{\mathrm{tot}}(\boldsymbol{q},\omega)$ without an external perturbation, indicating a self-sustaining wave. Since the time and spatial dependence of the density perturbation of the collective mode is
$n_{q}\exp\left(i\left[\boldsymbol{q}\cdot\boldsymbol{r} - \omega_{\boldsymbol{q}}t\right]\right)$, if $\mathrm{Im}(\omega_{\boldsymbol{q}})>0$ then the density perturbation grows exponentially (at least initially, until non-linear effects become apparent).

The dielectric function $\varepsilon(\boldsymbol{q},\omega)$ can be obtained at various levels of approximation. We use the long-wavelength limit of the random-phase approximation,\cite{mahan} which can be derived by solving the quantum-mechanical equations of motion of the charge carriers under the influence of a mean field electric potential, and Poisson's equation. The long-wavelength limit gives results that can also be derived using classical plasma physics, together with the relationship between the velocity and momentum; {\em i.e.}, in the case of parabolic band systems, $\boldsymbol{v}(\boldsymbol{p})$ $=$ $\boldsymbol{p}/{m}$, where $m$ is the effective mass of the carriers, and, in the case of graphene, $\boldsymbol{v}(\boldsymbol{p}) =  v_0\boldsymbol{p}/{\boldsymbol{|p|}}$, where $v_0$ is the speed of the electron in graphene.

We assume that our system is doped at a density $n_1$ of carriers, and there is no magnetic field. We also assume that the injected beam of carriers of density $n_2$ is peaked around a particular momentum $p_0$.  In the random phase approximation, the polarizability of the two charge carriers confined to the same space is equal to the sum of their individual polarizabilities, and hence the dielectric function has the form\cite{mahan}
\begin{equation}
\varepsilon(\boldsymbol{q},\omega) =  1 - V_c(q)[\Pi_1(\boldsymbol{q},\omega)+\Pi_2(\boldsymbol{q},\omega)]\ ,\label{eq:1}
\end{equation}
where $\Pi_1(\boldsymbol{q},\omega)$ and $\Pi_2(\boldsymbol{q},\omega)$ are the polarizability of the equilibrium carriers in the doped system and the injected carriers respectively, and $V_c(q)$  is the Fourier transform of the bare Coulomb interaction (statically screened by a background dielectric) with appropriate to the dimension of the system considered. We shall concentrate on long wavelength (small $q$) collective modes. In this limit the polarizabilities are given by
\begin{equation}
\Pi (\boldsymbol{q},\omega) = \int d\boldsymbol{p}\ \frac{\boldsymbol{q}\cdot\nabla f(\boldsymbol{p})}{\boldsymbol{q}\cdot\boldsymbol{v}-\omega}\ ,\label{eq:2}
\end{equation}
which is the small $q$ expansion of the random phase approximation.\cite{anal_cont} Here, $f(\boldsymbol{p})$ is the distribution function, with the normalization condition $n = \int d\boldsymbol{p}\; f(\boldsymbol{p})$, and $\boldsymbol{v}(\boldsymbol{p}) = \nabla_{\boldsymbol{p}} E$, where $E(\boldsymbol{p})$ is the band kinetic energy.

We assume that the doping carriers (subsystem 1) has an equilibrium distribution at zero temperature ({\em i.e.}, $f_1(\boldsymbol{p}) = $ constant for $\boldsymbol{|p|} < p_F$ and $f(\boldsymbol{p}) = 0$ for $\boldsymbol{|p|} > p_F$) which is also appropriate when the temperature is much less than the Fermi temperature.  Then in the small-$q$, non-zero frequency limit
\begin{eqnarray}
\Pi_1(\boldsymbol{q},\omega) = C\frac{n_{1}q^{2}}{\omega^{2}}\ ,\label{eq:3}
\end{eqnarray}
where $C$ is a constant which is dependent on the system dimension and energy dispersion $E(\boldsymbol{p})$.

For the injected carriers, we assume that the distribution is constant within a certain ``distance" (in momentum-space) of $\boldsymbol{p}_0$  and zero otherwise, which in the small $q$ limit can be approximated by $f_2(\boldsymbol{p}) = n_2\,\delta(\boldsymbol{p}-\boldsymbol{p}_0)$. The form of $\Pi_2(\boldsymbol{q},\omega)$ is dependent on the system dimension and the $E(\boldsymbol{p})$.   We also assume that $f_2(\boldsymbol{p})$ retains its $\delta$-function distribution over the length of the system.  This assumption is of course an approximation, because this is a non-equilibrium distribution that is subject to collisions that will thermalize it.  We argue in Sec.~\ref{sec:discussion} that for a certain experimentally relevant range of parameters, the thermalization is not significant over the relevant device length scales, justifying the approximation the injected electrons retain their $\delta$-function distribution throughout the system.

In the following sections we investigate the unstable modes which result from systems with an injected non-equilibrium beam of charged carriers. Before we present the results for graphene, we discuss the more familiar case of instability in three- and two-dimensional solid-state systems with parabolic energy band.  We then compare the dispersion relation, region and strength of instability of the parabolic band and graphene cases.

\section{Parabolic Bands}

In this section, we review the two-stream instability for parabolic bands where $E(\boldsymbol{p})$ $=$ $p^{2}/2m$, where  $m$ is the effective mass. The three-dimensional case has been well-studied by plasma physicists as it corresponds to a standard classical plasma where one charge distribution is streaming relative to another.\cite{chen_intro_plasma} The two-dimensional case which corresponds to injection of a non-equilibrium beam of electrons into a two-dimensional electron gas such as a doped quantum well has also been considered by various authors.\cite{robi67,krasheninnikov1980instabilities,crow85,hawr86,baks88,PhysRevB.39.6208,PhysRevB.43.14009,PhysRevB.48.9158,Zeba20122309}

For parabolic bands, the polarizability of the equilibrium distribution subsystem 1 in the non-zero $\omega$, small $q$ limit is\cite{pines1}
\begin{eqnarray}
\Pi_1(\boldsymbol{q},\omega) = \frac{n_{1}q^{2}}{m\omega^{2}}\ , \label{eq:4}
\end{eqnarray}
and, as shown in the appendix, the polarizability of the injected beam of charges is
\begin{eqnarray}
\Pi_2(\boldsymbol{q},\omega) = \frac{n_{2}q^{2}}{m(\omega-qv_0\cos\theta)^{2}}\ , \label{eq:5}
\end{eqnarray}
where $\theta$ is the angle between $\boldsymbol{q}$ and $\boldsymbol{v_0}$, the velocity of the injected carriers.

Using Eqs.~(\ref{eq:4}) and (\ref{eq:5}) in Eq.~(\ref{eq:1}) and setting $\varepsilon(\boldsymbol{q},\omega)$ equal to zero, gives the dispersion relation for three dimensional systems with parabolic band of
\begin{eqnarray}
1=\frac{4\pi e^{2}}{m\kappa}\left[\frac{n_{1}}{\omega^{2}}+\frac{n_{2}}{(\omega-qv_0\cos\theta)^{2}}\right]\ ,\label{eq:6}
\end{eqnarray}
and for two dimensional systems of
\begin{eqnarray}
1=\frac{2\pi e^{2}q}{m\kappa}\left[\frac{n_{1}}{\omega^{2}}+\frac{n_{2}}{(\omega-qv_0\cos\theta)^{2}}\right]\ ,\label{eq:7}
\end{eqnarray}
where $\kappa$ is the dielectric constant of the background material.
We note that these dispersion relations can also be obtained using Newton's equations of motion and Poisson's equation, which is the standard approach that is used in classical plasma physics.\cite{chen_intro_plasma} The random-phase-approximation formalism that we have used, however, can be generalized to include quantum effects (which are important when $q$  is not small), while the classical plasma physics method cannot.

Comparing Eqs.~(\ref{eq:6}) and (\ref{eq:7}), there is an additional factor of $q/2$ on the right hand side of Eq.~(\ref{eq:7}), which arises from the different forms of the Fourier transform of the Coulomb interaction in two and three dimensions.
This factor of $q/2$  results in significant differences between the instabilities in three and two dimensions.

In Eqs.~(\ref{eq:6}) and (\ref{eq:7}), the equilibrium value of the equilibrium charge density $n_1$  is not necessarily equal to that of the injected charge carriers $n_2$.  Typically, in standard neutral plasmas, the densities are equal, since $n_1$ usually corresponds to the density of the positive ions and $n_2$  to the density of electrons.  Since we are dealing with solid state systems, we assume that there is a uniform background charge that compensates for any net charge of the mobile charge carriers including injected charge.

Multiplying Eqs.~(\ref{eq:6}) and (\ref{eq:7}) by their respective denominators of the terms on the right hand side gives a fourth order polynomial equation for $\omega$, so four roots are expected. The instability of the plasma wave is determined by whether the roots are all completely real or only two are completely real and two have non-zero imaginary parts. If only two roots are completely real, the remaining two roots appear as a complex conjugate pair. The root with a positive imaginary value of $\omega$ ({\em i.e.}, $\mathrm{Im}(\omega)>0$ ) indicates the exponentially growing wave, as can easily be seen by inserting a complex $\omega$  into the time-dependence of the wave, $\exp(-i\omega t)$, which results in a factor of $\exp[\mathrm{Im}(\omega) t]$. Thus, $\mathrm{Im}(\omega)$ gives the growth rate of the unstable wave.\cite{chen_intro_plasma}

The solutions of the roots of Eqs. (\ref{eq:6}) and (\ref{eq:7}) were obtained numerically using Matlab.   We assumed the ratio $n_1/n_2 = 10$, where  $n_1 = 10^{18}\,\mathrm{cm}^{-3}$ and $n_1=10^{12}\,\mathrm{cm}^{-2}$ for the three- and two-dimensional cases, respectively.

Fig.~1(a) and Fig.~2(a) show the imaginary part of the complex root of Eqs.~(\ref{eq:6}) and (\ref{eq:7}), respectively. It can be seen that for both three-and two-dimensional systems; the instabilities occur at low wavenumber and disappear above a certain cutoff $q$.  In three-dimensional systems, the peak value of $\mathrm{Im}(\omega)$ is the same for any given angle, but the peak wavenumber shifts to higher $q$ as the angle $\theta$ between $\boldsymbol{q}$  and $\boldsymbol{p}_0$ increases. The reason for this evident when the dispersion relation Eq.~(\ref{eq:6}) is examined. The only dependence on the dispersion on $q$ is in the denominator of the second term on the right, where $q$ is multiplied by $\cos\theta$. Therefore, the effect of changing the angle $\theta$ is to ``renormalize" the value of $q$  for $\theta=0^{\circ}$ to $q\cos\theta$  for $\theta\neq 0^{\circ}$. In the two dimensional case, in addition to a shift in the position of the peak of the growth rate $\mathrm{Im}(\omega)$ towards increasing value of $q$ with increasing angle $\theta$ that is seen in the three-dimensional case, the magnitude of the peak also increases. This is due to the additional factor of $q$ on the right-hand side of the dispersion relation. Figures (b) and (c) of both Figs.~1 and 2 show the real and imaginary parts of the dispersion relations for a representative small ($10^{\circ}$) and moderately large ($45^{\circ}$) angle of $\boldsymbol{q}$ with respect to $\boldsymbol{p}_0$. There are four modes, two of which have frequencies that are always real, which correspond to the standard plasmon propagating in opposite directions, and two of which come in complex conjugate pairs below a certain $q$ (which depends on the angle).

These results for the growth rates are reproduced as surface and contour plots in Figs.~3 and 4 for three dimensional and two dimensional parabolic band system respectively, where it is assumed that the momentum of the injected charge carriers is in the positive $x$-direction. We see easily from these plots that in the three-dimensional case the instability depends only on the component of $q_x$ (the component of $\boldsymbol{q}$ in the direction of charge carrier injection) and is independent of $q_y$ (which, by symmetry is equivalent to any component of $\boldsymbol{q}$ that is perpendicular to the x-direction). This is easily explained by inspection of the dispersion relation Eq.~(\ref{eq:6}), where the only dependence on the wavenumber $q$ occurs as $q\cos\theta$, which is equal to $q_x$. In the two-dimensional case the growth rate is also dependent of $q_y$, because of the presence of the additional factor of $q=(q_{x}^2+q_{y}^2)^{\frac12}$ in the dispersion relation Eq.~(\ref{eq:7}), as compared to Eq.~(\ref{eq:6}) for the three-dimensional case.

The plots presented in Figs.~3 and 4 also explain what appears to be a contradictory result in Figs.~1(a) and 2(a). The range of wavevectors for unstable waves grows as $\theta$ increases from $0^{\circ}$ to $90^{\circ}$ but appears to disappear abruptly and completely when $\theta = 90^{\circ}$. Figs.~3 and 4  show that there is no discontinuity when $\theta \rightarrow 90^{\circ}$.  What has happened in Fig.~1 and 2 for $\theta =  90^{\circ}$  is that the peak of the instability has been pushed to $q$ $\rightarrow$ $\infty$, therefore it does not appear on the plot.  All that is visible is the $q \rightarrow 0$  dependence of $\mathrm{Im}(\omega)$, which vanishes. We remind the reader that we have used the small $q$ approximation and therefore the results presented here are not reliable at large $q$.


\section{Graphene}

We now investigate the instability when the energy dispersion is $E(p)=pv_0$  in a two-dimensional system as in the case of graphene, where the momentum is taken with respect to the $K$ and $K'$ points in the Brillouin zone. The velocity is taken to be $v_0 = 1.0\times10^{8}\,$cm/s as in graphene.  We take the density of the $ n_1=10^{12}\,\mathrm{cm}^{-2}$, which gives a Fermi energy of $\varepsilon_F = \hbar v_0 \sqrt{\pi n_1} = 120\,\mathrm{meV}$, and the ratio of the density of the equilibrium electrons in the system to the injected electrons to be $n_1/n_2 = 10$.  We assume that the electrons are injected into the system at energy of $E=10\,$meV above the Fermi energy.

For the equilibrium electrons at zero temperature, where the distribution function is constant for $\boldsymbol{|p|}<p_F$ and zero for $\boldsymbol{|p|}> p_F$, the polarizability for the equilibrium electrons in the small $q$ limit is\cite{PhysRevB.75.205418}
\begin{eqnarray}
\Pi_1(\boldsymbol{q},\omega) = \frac{n_{1}v_0q^{2}}{p_F\omega^{2}}\,.\label{eq:8}
\end{eqnarray}
and, as shown in the appendix, the polarizability for a distribution of density $n_2$ that is peaked around $\boldsymbol{p}_0$ which makes an angle $\theta$ with respect to $\boldsymbol{q}$ is
\begin{eqnarray}
\Pi_2(\boldsymbol{q},\omega) = \frac{n_{2}v_0 q^{2}\sin^2{\theta}}{p_0(\omega-qv_0\cos\theta)^{2}}\ .\label{eq:9}
\end{eqnarray}
Thus setting the dielectric function equal to zero in Eq.~(\ref{eq:1}) gives the dispersion relation
\begin{eqnarray}
1=\frac{2\pi e^{2}v_0q}{\kappa}\left[\frac{n_{1}}{\omega^{2}}\frac{1}{p_F}+\frac{n_{2}}{(\omega-qv_0\cos\theta)^{2}}\frac{\sin^2{\theta}}{p_0}\right]\ ,\label{eq:10}
\end{eqnarray}
where $\kappa$ is the dielectric constant in the 2-dimensional layer.  Eq.~(\ref{eq:10}) can be rewritten in the form
\begin{eqnarray}
1=\frac{\epsilon}{z^{2}}+\frac{1}{(z-\lambda)^{2}}\ , \label{eq:11}
\end{eqnarray}
where
\begin{subequations}
\begin{eqnarray}
\epsilon=\frac{n_{1}p_0}{n_{2}p_F\sin^2{\theta}}\ ,\label{eq:12a}
\end{eqnarray}
\begin{eqnarray}
z=\omega\sqrt\frac{\kappa p_0}{2\pi e^{2}v_0n_{2}q}\frac{1}{\sin\theta}=\frac{\omega}{\omega^{*}(q)}\ ,\label{eq:12b}
\end{eqnarray}
\begin{eqnarray}
\lambda=\sqrt\frac{\kappa p_0v_0q}{2\pi e^{2}n_{2}}\cot\theta=\frac{qv_0\cos\theta}{\omega^{*}(q)}\ ,\label{eq:12c}
\end{eqnarray}
and where
\begin{eqnarray}
\omega^{*}(q)=\sqrt\frac{2\pi e^{2}v_0n_{2}q}{\kappa p_0}\sin\theta\ ,\label{eq:12d}
\end{eqnarray}
\end{subequations}
is the long wavelength plasmon dispersion for a density of carrier $n_2$ that is injected into the system with momentum $\boldsymbol{p}_0$. This results in a quartic equation in $z$
\begin{eqnarray}
z^{4}-2\lambda z^{3}+(\lambda^{2}-\epsilon-1)z^{2}+2\epsilon\lambda z-\lambda^{2}\epsilon=0\ .\label{eq:13}
\end{eqnarray}

Solving analytically, two of the four roots have non-zero imaginary parts when\cite{bellon}
$ 
\lambda<(1+\epsilon^{1/3})^{2/3}. 
$ 
Since the coefficients of the quartic equations are real, the complex roots come in complex conjugate pairs. In Fig.~\ref{fig:5}, we plot the imaginary part of the complex root of Eq. (\ref{eq:13}) (i.e. $\mathrm{Im}(z)$ as a function of $\lambda$ and $\epsilon$).
In order to obtain the instability growth rates as a function of $\boldsymbol{q}=(q_{x}, q_{y})$ for certain material parameters, one first obtains $\lambda$ and $\epsilon$ for that $\boldsymbol{q}$ and the material parameters. The $\mathrm{Im}[z(\lambda, \epsilon)]$ is obtained from Eq.~(\ref{eq:11}), and the growth rate can  then be obtained from Eq.~(\ref{eq:12b}); {\em i.e.}, $\mathrm{Im}(\omega) = \omega^{*}(q)\mathrm{Im}(z)$ where $\omega^{*}(q)$ is given by Eq.~(\ref{eq:12d}). The results for certain parameters are shown in Figs.~6 and 7. As in the case of the instability in the parabolic-band case in two dimensions, for a given $q_x$, the growth rate increases with increasing $q_y$, and the instability disappears for $\boldsymbol{q}$ perpendicular to the momentum of the injected carriers, $\boldsymbol{p}_0$. However, there is an important qualitative difference between the parabolic band and graphene cases. For parabolic bands, instabilities occur (up to a certain wave-number) for $\boldsymbol{q}$ along the $x$-axis; that is, parallel to the direction of injection of the carriers. In the case of graphene, there is {\em no} instability along the $x$-axis.  Mathematically, this is because of the presence of the $\sin^{2}\theta$ term in the polarizability $\Pi_{2}(\boldsymbol{q}, \omega)$ for the injected carriers graphene, which is absent in the equivalent polarizability in the parabolic-band case. As in Figs.1 (b), (c) and 2 (b), (c), the Fig. 6 (b) and (c) show the real and imaginary parts of the dispersion relations for a small angle ($10^{\circ}$) and a moderately large angle ($45^{\circ}$) of $\boldsymbol{q}$ with respect to $\boldsymbol{p}_0$ respectively. The major difference between the graphene and the parabolic-band cases occurs for small angles. Note that in the graphene case, for the $10^{\circ}$ graph, the scale of the wavevector and growth rates are roughly an order of magnitude smaller than for the $45^{\circ}$ graph.  In the next section, we discuss the reason for this significant qualitative difference between the parabolic-band and graphene cases.
\section{Discussion and Conclusion}\label{sec:discussion}

The absence of instabilities in modes with $\boldsymbol{q}$ in the same direction of the injected carriers $\boldsymbol{p}_0$ in graphene is due to the linear energy dispersion. To understand why this is so, consider the mechanism for the formation of unstable modes. When there is a small spatial fluctuation in the charge density (caused by, for example, a thermal fluctuation) the resultant electric potential perturbation due to this fluctuation causes a change in the motion of the charge carriers. In systems that are in equilibrium, the net effect of the change in the motion of the charge carriers is to reduce the magnitude of the spatial fluctuation in charge. However, under certain non-equilibrium situations, as in the case of two-stream instabilities, a spatial fluctuation in the charge density causes a change in the motion of the charge carriers which tends to increase the fluctuation, resulting in a charge fluctuation that grows (initially) exponentially. Note that in order for this unstable feedback loop to occur, the spatial fluctuation in the charge density has to change the motion of the charge carriers.

In the case of a one-dimensional system where the energy dispersion is linear, the carriers in an injected beam centered on momentum $\boldsymbol{p}_0$, say in the positive direction, all have the same velocity $v_0$. As the velocity of the particles is independent of the momentum, any change in the momentum of the particles due to the forces caused by the electric potential caused by a spatial fluctuation in the charge will $not$ change the motion of the particles. Thus, the unstable feedback loop never occurs and there is no instability.

This explains why there is no instability in two-dimensional graphene system when $\boldsymbol{q}$ is in the same direction of the injected carriers $\boldsymbol{p}_0$, since this situation mirrors that of the one-dimensional system.  But since the speed of particles is the same for any momentum, why does this argument not work when $\boldsymbol{q}$ is not in the same direction as $\boldsymbol{p}_0$? As can be seen from the results shown in Figs.~6 and 7, there are robust regions of instability for $\boldsymbol{q}$ not parallel to $\boldsymbol{p}_0$.

This is because while the $speed$ of the particles is constant, their {\em velocities} (which also include the direction of motion) depend on the momentum. Therefore, when $\boldsymbol{q}$ is not parallel to $\boldsymbol{p}_0$, the density perturbation of the wave produces a force which affects the direction of the particles in the beam centered at $\boldsymbol{p}_0$, which affects the velocity of the particles. This allows the unstable feedback loop to occur, resulting in (initially) exponentially growing waves.

We note that in this paper, we have made several simplifying assumptions.
We have used a small wavevector approximation, which will break down when at wavevectors that are on the order of the smallest characteristic wavevectors of the distributions of the electrons. Therefore, the unbounded increase in the instability growth rates with increase in the magnitude of the wavevectors that are seen in the two-dimensional instabilities are expected to be an artifact of the small wavevector approximation. We have also not taken into account the effect of the lattice scattering on the distributions and the instability rates.  The lattice scattering will tend to cause diffusive behavior in the small wavevector regime, which will tend to suppress the instability growth rates.

Experimentally, one possible method for obtaining the non-equilibrium distribution of injected carriers in doped graphene that is required for the occurrence of the two-stream instability is to use a planar tunnel junction attached to an edge of a doped graphene flake.\cite{contacts}  The detection of the unstable waves could be achieved by placing, at the opposite side of the graphene flake from the injection end, several individually contacted drains which are at different angles with respect the direction of the injected electrons.  There should be a variation in the magnitudes of currents detected by drains, with the largest enhancement occurring for gates that are placed at roughly $45^{\circ}$ with respect to the direction of injection of the electrons, since the growth rates of the unstable waves in those directions are the largest.  Since the frequencies of the unstable waves range up to the terahertz regime, these waves may be beyond the usual detection limit of standard electronic devices.  In this case, it may be possible to detect the waves via a grating coupler in the drain region, which is designed to couple to the terahertz plasma waves and emit terahertz radiation,\cite{emitTHz} which can then be detected.\cite{detectTHz}  In fact, using this technique, this instability might potentially be used as a source of terahertz electromagnetic waves.

The possibility of observing two-stream instabilities in graphene are enhanced over standard parabolic-band two-dimensional electron gases.  In parabolic band systems, in order to obtain the conditions for a two-stream instability, the non-equilibrium electrons must be injected at high energy to obtain the necessary speed.  This makes them susceptible to the strong inelastic scattering processes that typically occur at high energies (such as optic-phonon and electron-electron scattering), which quickly degrade the current.   In graphene, the speed of the electrons in is independent of the energy, and therefore there is no need to inject the carriers at high energies in order to obtain the conditions necessary for a two-stream instability.

This fact supports our approximation that the non-equilibrium distribution function maintains its form ({\em i.e.}, it is strongly peaked at the injection energy) as it passes through the system.  The main contribution to the thermalization is the electron--electron scattering of the injected carriers with the equilibrium carriers that are already in the sample.   The single-particle lifetime of an electron in graphene due to electron--electron scattering with a fermi sea of electrons at zero temperature $\tau_{\mathrm{ee}}$ is given by\cite{PhysRevB.75.121406} $\tau_{\mathrm{ee}}^{-1} = - 2\, \mathrm{Im}[\Sigma_{\mathrm{ret}}] = \frac{\xi^2}{4\pi \varepsilon_F} \left[\log\left(\frac{8\varepsilon_F}{\xi}\right) - \frac12\right] $, where $\Sigma_{\mathrm{ret}}$ is the on-shell retarded self-energy, $\varepsilon_F$ is the Fermi energy and $\xi$ is the energy of the electron relative to $\varepsilon_F$.  For $\xi = 10\,\mathrm{meV}$ and $\varepsilon_F = 120\,\mathrm{meV}$, corresponding to the parameters used in Figs.~(\ref{fig:6}) and (\ref{fig:7}), this gives $\tau_{\mathrm{ee}} = 2.4\,\mathrm{ps}$, which translates to a mean free path of $2.4\,\mu$m.  This is considerably larger than the wavelengths a good portion of the unstable modes.  Furthermore, the $\tau_{\mathrm{ee}}$ represents the single-particle lifetime, which takes into account {\em all} scattering out of a given momentum state, regardless of how small the momentum transfer is.  Due to the long-ranged nature of the Coulomb scattering, the small-momentum scattering events dominate, but these events are very inefficient at degrading the non-equilibrium distribution function.  A more appropriate measure of the rate of thermalization of the distribution is the transport lifetime, which weights scattering events by the momentum loss of the electron.  We are not aware of any calculations of the transport relaxation rate due to electron--electron scattering in graphene, but if we assume that charged-impurity scattering rates are an acceptable proxy for electron--electron scattering, then the transport lifetime can be over an order of magnitude larger than the single-particle lifetimes.\cite{PhysRevB.77.195412}    This furthers bolsters the validity of our approximation of a constant distribution function for the injected electrons across the device.\cite{double_layer}

Finally, we contrast the two-stream instability described here with the more well-known Dyakonov-Shur (DS) instability,\cite{PhysRevLett.71.2465} which has been studied experimentally in field-effect transistors\cite{DS_exp_review} and theoretically in graphene.\cite{PhysRevB.88.205426,PhysRevB.93.245408,doi:10.1117/12.2227438}   The DS instability occurs in two-dimensional electron gases, such as field effect transistors, which have a certain range of drift velocities and are subject to boundary conditions of constant total current at one end and constant the electric potential at the other.   This leads under appropriate conditions to the amplitudes of density perturbations inside the two-dimensional electron gas being enhanced when they reflect off the boundaries, resulting in unstable waves.  The two-stream instability on the other hand is a bulk effect, and the boundaries of the sample do not drive the effect.   Another major difference is that the DS instability is based on a hydrodynamic theory, which assumes that the electron--electron scattering dominates to the extent that the distribution function is locally a drifted thermalized  distribution.  This is the opposite limit from the two-stream instability, where it is crucial that the electron--electron scattering is at a low enough level to prevent thermalization of the non-equilibrium injected electrons.

In conclusion, we have studied two-stream instabilities in two-dimensional systems with linear energy dispersions, such as in graphene, in which a beam of electrons is injected into a doped system. The range of wavevectors in which instabilities occur in the case of graphene is qualitatively different from that of conventional three- and two-dimensional parabolic-band systems. In particular, there is a complete absence of instabilities in waves with wavevectors that are parallel to the direction of injection of the non-equilibrium electrons.
\begin{acknowledgments}
CMA would like to acknowledge Dr.~Jutta Luettmer-Strathmann for her assistance in using Matlab.  BYKH acknowledges support by the FiDiPro program at Aalto University and a Faculty Summer Fellowship at the University of Akron.  The Center for Nanostructured Graphene (CNG) is sponsored by the Danish National Research Foundation, Project DNRF103.
\end{acknowledgments}

\appendix*
\section{Derivations of Eqs.~(\ref{eq:5}) and (\ref{eq:9})}

The expressions for the polarizability in  Eqs.~(\ref{eq:5}) and (\ref{eq:9}) are derived from substituting the sharply peaked distribution $n(\mathbf p) = n_2\delta(\mathbf p-\mathbf p_0)$ into Eq.~(\ref{eq:2}).  Let $\mathbf q = q\hat{\mathbf x}$, and $\mathbf p_0$ make and angle of $\theta$ with respect to the $x$-axis, so that $p_{0,x} = p_0 \cos\theta$ and $\vert \mathbf p_{0,\perp}\vert = p_0 \sin\theta$, where $\mathbf p_\perp$ are the components of $\mathbf p$ that are perpendicular to the $x$-axis.  Then,
\begin{align}
\Pi_2(\mathbf q,\omega) &= n_2 \int_{-\infty}^\infty dp_x \int_{\mathrm{all\ }\mathbf p_\perp}\!\!\!\! d\mathbf p_\perp \frac{q {\displaystyle\frac{\partial}{\partial p_x}} \delta(p_x-p_{0,x})\ \delta(\mathbf p_\perp - \mathbf p_{0,\perp})}{q v_x(p_x,\mathbf p_\perp) - \omega}\nonumber\\
&= n_2 \int_{-\infty}^\infty dp_x  \frac{q {\displaystyle\frac{\partial}{\partial p_x}} \delta(p_x-p_{x,0})}{q v_x(p_x,p_{0,y}) - \omega}\nonumber\\
&= n_2 q^2\int_{-\infty}^\infty dp_x\ \delta(p_x - p_{0,x})\; \frac{\partial v_x(p_x,p_{0,y})}{\partial p_x}\; \frac1{(q v_x(p_x,p_{0,y}) - \omega)^2}\nonumber\\
&= n_2 q^2 \frac1{\bigl(q\,v_x(\mathbf p_0) - \omega\bigr)^2} \left[\frac{\partial v_x(p_x,\mathbf p_{0,\perp})}{\partial p_x}\right]_{p_x=p_{0,x}}. \label{eq:A1}
\end{align}
where in the above we have integrated by parts to obtain the third equality.

For parabolic bands, $v_x= p_x/m$, so that $\displaystyle\frac{\partial v_x}{\partial p_x} = \frac1m$.  Substituting this into Eq.~(\ref{eq:A1}) gives Eq.~(\ref{eq:5}).  For a linear bands in 2 and 3 dimensions, $\displaystyle v_x = v_0 \frac{p_x}{\sqrt{p_x^2 + p_\perp^2}}$, and therefore $\displaystyle\frac{\partial v_x}{\partial p_x} = v_0\frac{p_\perp^2}{\vert\mathbf p\vert^3}$.  Substituting this into Eq.~(\ref{eq:A1}), and using $\vert \mathbf p_{0,\perp}\vert = p_0 \sin\theta$, gives Eq. (\ref{eq:9}).

Note that for linear bands in one-dimension, $v_x$ is constant and therefore $\displaystyle\frac{\partial v_x}{\partial p_x} = 0$.  (This reflects the fact that in one-dimensional systems with a linear band, the motion of the particles does not depend on forces it experiences.)  Therefore, two stream instabilities do not occur in one-dimensional systems with linear bands.

\newpage
\begin{center}
\underline{\bf Figures}
\end{center}

\begin{figure}[hb]  
\includegraphics[width=\textwidth,height=8.5cm]{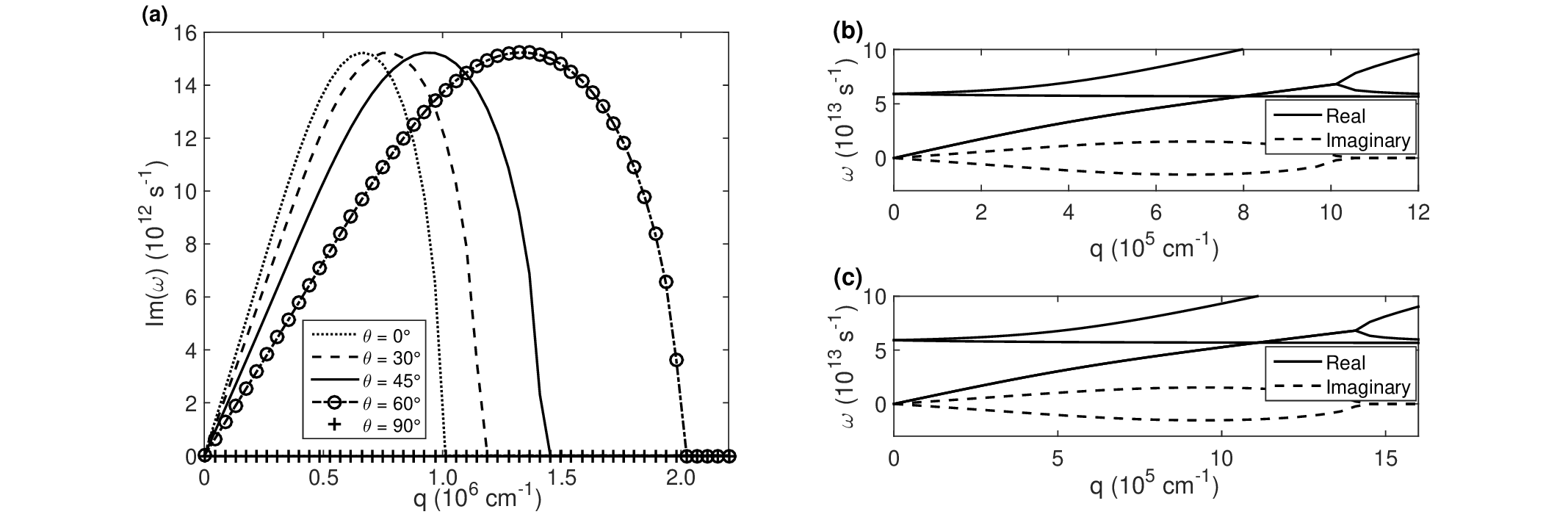}
\caption{\label{fig:1}(a) Imaginary part of $\omega$ (the growth rate) as a function of wave number for different angles $\theta$ between wavevector $\boldsymbol{q}$ and injected carrier momentum $\boldsymbol{p}_0$, for a three-dimensional parabolic band system.  The mass is taken to be the bare electron mass, $\kappa=1$, $n_1 = 10^{18}\,\mathrm{cm}^{-3}$, $n_1/n_2 = 10$ and the energy of the injected electrons is $E =$ 10\,meV above the Fermi energy.  Figs.~(b) and (c) show the dispersion relation for angles $\theta=10^{\circ}$ and $\theta= 45^{\circ}$ respectively.  The solid lines are the real part of $\vert\omega\vert$ and the dashed lines are the imaginary part of the modes that split at around $1.0\times 10^6\,\mathrm{cm}^{-1}$ and $1.5\times 10^6\,\mathrm{cm}^{-1}$ in (b) and (c), respectively. }
\end{figure}  
\bigskip\bigskip

\begin{figure*}
\includegraphics[width=\textwidth,height=8.5cm]{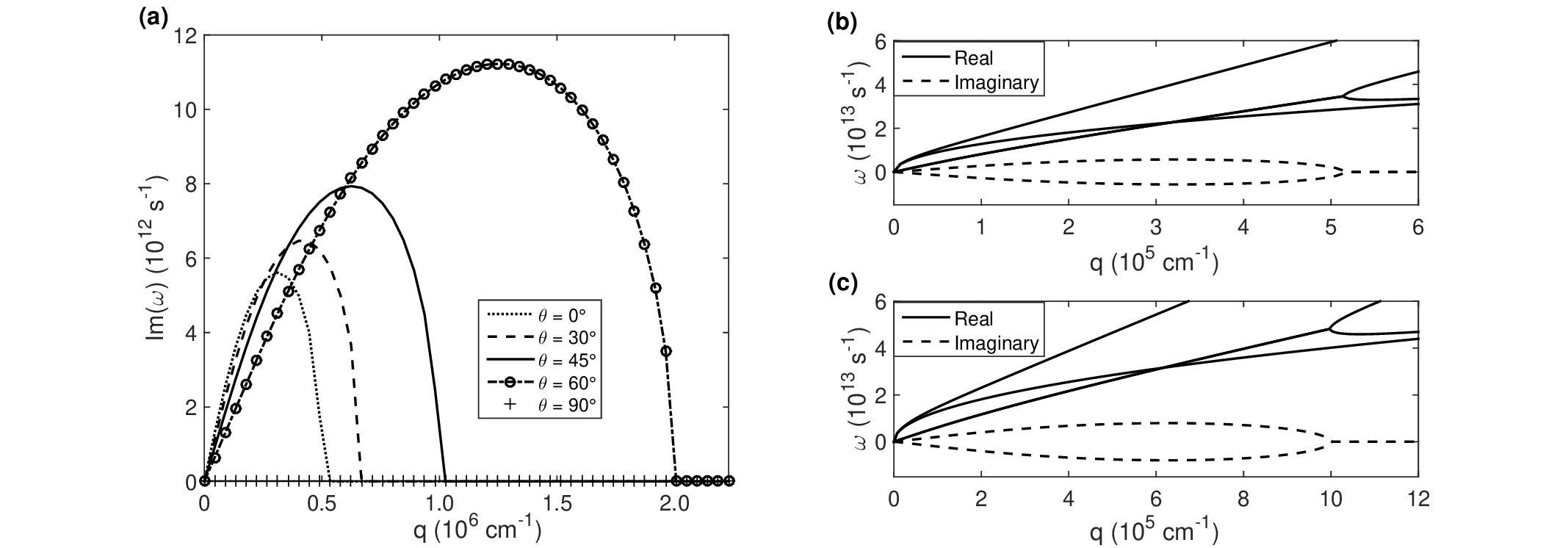}
\caption{\label{fig:2}Same as in Fig.~1 for a two-dimensional parabolic-band system with $n_1 = 10^{12}\,\mathrm{cm}^{-2}$, and other parameters unchanged.}
\bigskip\bigskip
\end{figure*}

\begin{figure*}
\includegraphics[width=\textwidth,height=6cm]{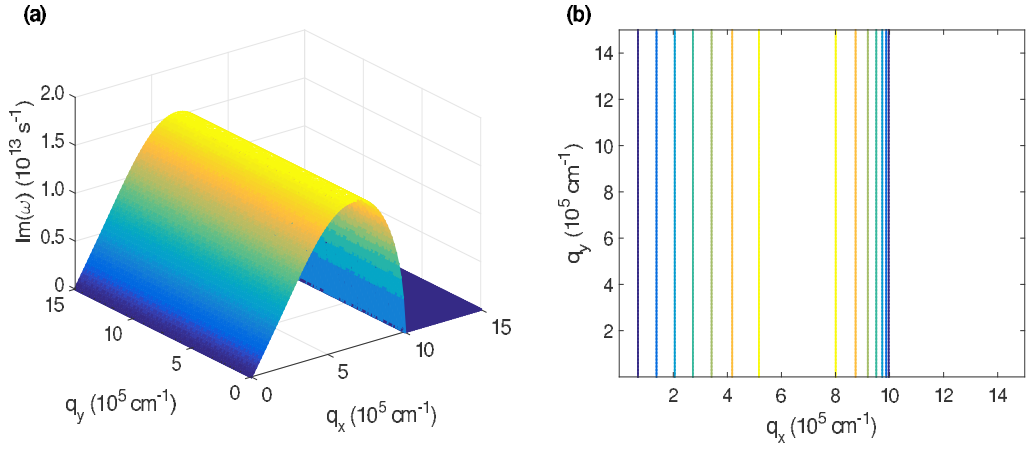}
\caption{\label{fig:3} Growth rates of the unstable plasmons ($\mathrm{Im}[\omega]$) as a function of $\boldsymbol{q}$ for a three-dimensional parabolic-band system, shown as (a) a surface plot and (b) a contour plot.  The parameters used are the same as Fig.~1. The dark flat region in (a), and correspondingly the region to the right of $q_x = 1.0\times 10^6\,\mathrm{cm}^{-1}$ in (b), is where $\mathrm{Im}(\omega) = 0$; {\em i.e.}, stable waves.  In (b), the lowest contour lines are at $2\times 10^{12}\,\mathrm{s}^{-1}$ and the difference between successive contour lines is $2\times 10^{12}\,\mathrm{s}^{-1}$.}
\end{figure*}

\bigskip\bigskip

\begin{figure*}
\includegraphics[width=\textwidth,height=6cm]{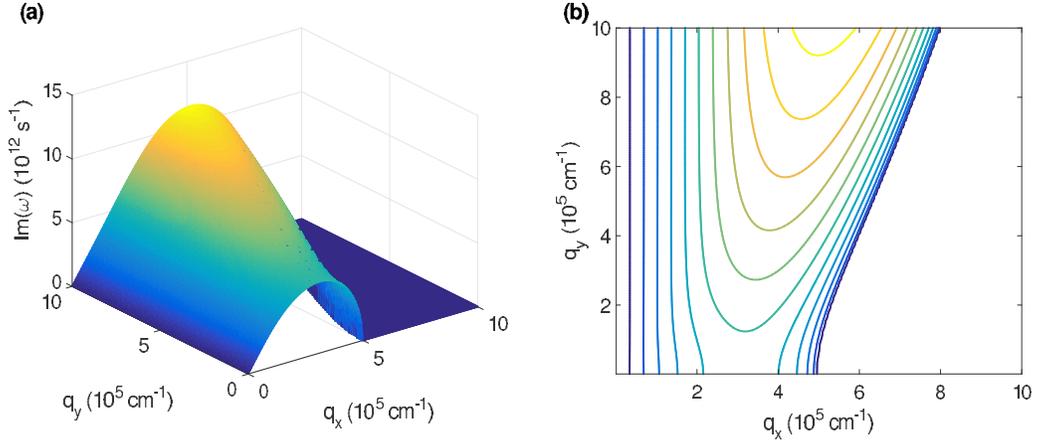}
\caption{\label{fig:4}Same as in Fig.~3 for a two-dimensional parabolic-band system with $n_1 = 10^{12}\,\mathrm{cm}^{-2}$, and other parameters unchanged. In (b), the lowest contour lines are at $1\times 10^{12}\,\mathrm{s}^{-1}$ and the difference between successive contour lines is $1\times 10^{12}\,\mathrm{s}^{-1}$.}
\end{figure*}

\bigskip\bigskip

\begin{figure*}
\includegraphics[width=\textwidth,height=6cm]{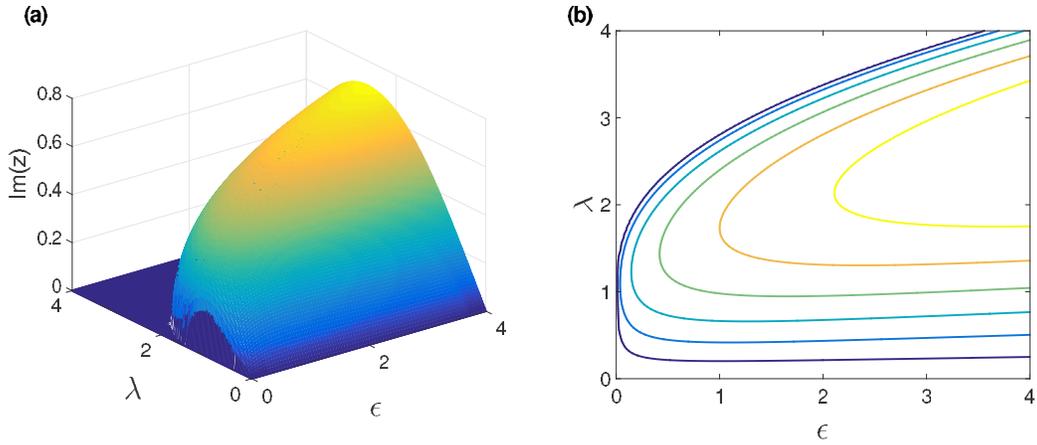}
\caption{\label{fig:5} (a) Surface plot of $\mathrm{Im}(z)$ as a function of mathematical parameters $\lambda$ and $\epsilon$, where $z$ is the complex root of Eq.(14); (b) corresponds the counter plot of (a), where the lowest contour line is at $0.1$ and the difference between successive contour lines is $0.1$.}
\end{figure*}
\bigskip\bigskip


\begin{figure*}
\includegraphics[width=\textwidth,height=8.5cm]{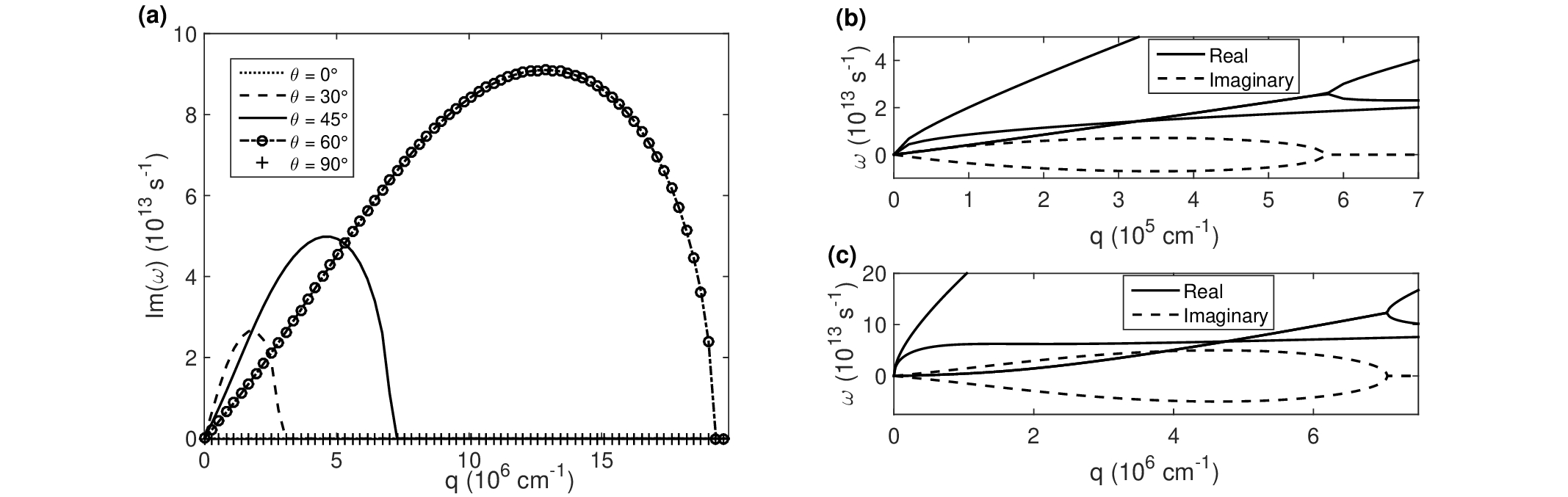}
\caption{\label{fig:6}(a) Imaginary part of $\omega$ (the growth rate) as a function of wave number for different angles $\theta$ between wavevector $\boldsymbol{q}$ and injected carrier momentum $\boldsymbol{p}_0$, for a linear-band system such as graphene. Here, $v_0=1.0\times10^{8}$ cm/s, $\kappa=3$ (as for graphene on a BN substrate), $n_1 = 1.0\times 10^{12}\,\mathrm{cm}^{-2}$ and $n_1/n_2=10$ and the energy of the injected electrons is $E =$ 10\,meV above the Fermi energy. For both $\theta = 0^{\circ}$ and $90^{\circ}$,  the imaginary part of $\omega$ is zero.   Figs.~(b) and (c) show the dispersion relations for angle $\theta$ $=$ $10^{\circ}$ and $\theta$ $=$ $45^{\circ}$ respectively; as in Figs.~1 and 2, the solid lines are the real part and the dashed lines are the imaginary part of the mode that splits.}
\end{figure*}
\bigskip\bigskip

\begin{figure*}
\includegraphics[width=\textwidth,height=6cm]{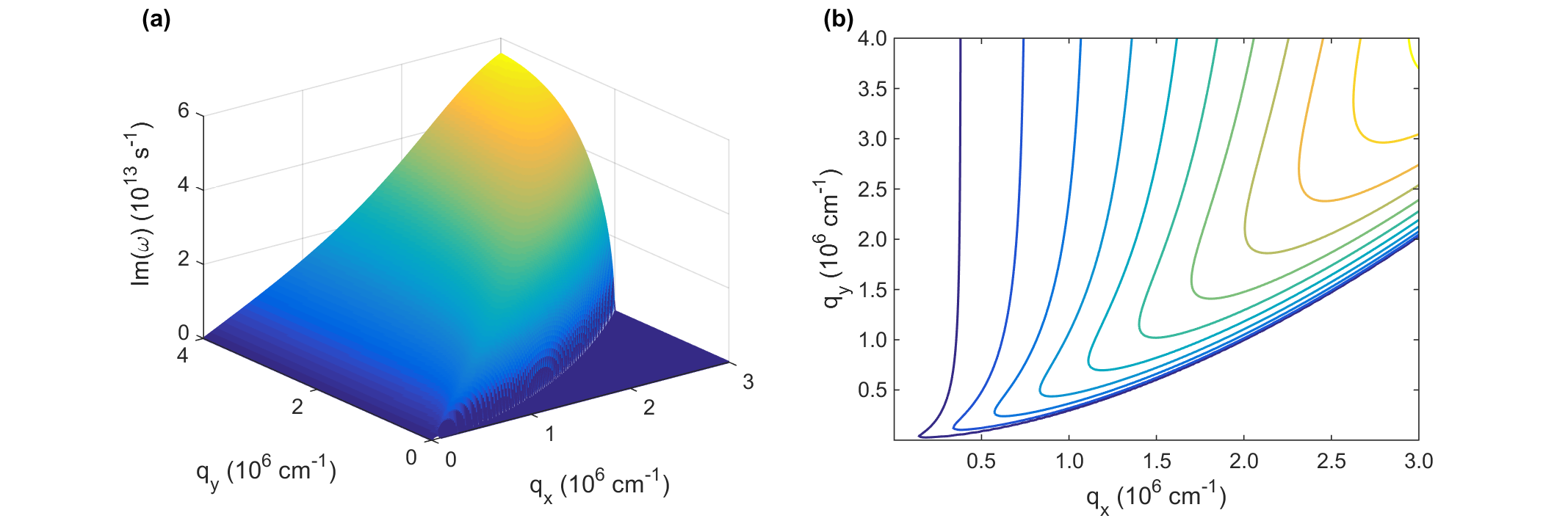}
\caption{\label{fig:7}(a) Surface plot and (b) contour plot for $\mathrm{Im}(\omega)$ as a function of $\boldsymbol{q}$ for graphene, with the same parameters as in Fig.~6. In (b), the lowest contour line is at $5\times 10^{12}\,\mathrm{s}^{-1}$ and the difference between successive contour lines is $5\times 10^{12}\,\mathrm{s}^{-1}$.}
\end{figure*}


\begin{thebibliography}{33}
\bibitem[{dou()}]{double_layer}
\bibinfo{note}{If necessary, alternative device configurations could be used to
  suppress the electron--electron scattering; {\em e.g.}, a double-layer
  graphene system with negligible tunneling between the layers (as used in
  Coulomb drag experiments) where the extrinsic electrons are in one layer and
  the electrons are injected in the other layer. In this case, the Coulomb
  interaction between the layers is reduced by a factor of $\exp(-qd)$, where
  $d$ is the separation between the layers. A distance can be chosen so that
  the high-momentum transfer interlayer Coulomb scattering is suppressed, but
  the longer wavelength unstable plasmons are not significantly affected.}
\end{thebibliography}
\end{document}